\documentclass[aps,amsmath,twocolumn,10pt,prl]{revtex4-1}
\usepackage{docs}%
\usepackage{bm}%
\usepackage[colorlinks=true,linkcolor=blue]{hyperref}%
\usepackage{graphicx}
\usepackage{dcolumn}
\expandafter\ifx\csname package@font\endcsname\relax\else
 \expandafter\expandafter
 \expandafter\usepackage
 \expandafter\expandafter
 \expandafter{\csname package@font\endcsname}%
\fi
\begin{document}

\title{Size dependent rearrangements in monometallic clusters}%

\author{K. Rossi, L. Pavan, Y. Soon and F. Baletto}%
\email{francesca.baletto@kcl.ac.uk}
\affiliation{Physics Department, King's College London, WC2R 2LS, UK}%

\date{July 2015}%
\revised{October 2015}%
\begin{abstract}
Morphology and its stability are essential features to address physicochemical properties of metallic nanoparticles. By means of Molecular Dynamics based simulations we show a complex dependence on the size and material of common structural mechanisms taking place in mono-metallic nanoparticles at icosahedral magic sizes.
We show that the well known Lipscomb's Diamond-Square-Diamond mechanisms, single-step screw dislocation motions of the whole cluster, take place only below a given size which is material dependent.
Above that size, layer-by-layer dislocations and/or surface peeling are likely to happen, leading to low symmetry defected motifs.
The material dependence of this critical size is similar to the crossover sizes among structural motifs, based on the ration between the bulk modulus and atomic cohesive energy.
\end{abstract}
\maketitle

Investigating structural transition in metallic nanoparticles is paramount to nurture their large scale application in a wide range of fields such as catalysis, optics, biomedicine and others.\cite{Nørskov2009, Stockman2015, Dykman2012}  
Morphology, together with size and composition, determines nanocluster's chemophysical properties,\cite{Calle-Vallejo2015b,  Asara2016a, DiPaola2016} establishing a profound link between structural stability and reliable performance.
Thus the lifetime of a nanoarchitecture has to be addressed to predict its strength against ageing.
Numerical and analytical methods are nowadays able to predict favourable morphologies and rationalize them in terms of the inter-atomic potential stickyness, edge and surface energy.\cite{Baletto2002f, a, b}
Computational techniques can also investigate kinetic and entropic trapping during the nucleation or formation process which may lock the system in a structure different from the minimum energy one.\cite{Baletto2005, Baletto2004b} 
Understanding whether or not clusters trapped in metastable configuration can rearrange towards more stable ones is a fundamental step in order to weight the kinetic's and thermal's contributions determining the polidispersivity as commonly detected in experimental populations distributions.\cite{Barnard, Barnard2012a, Barnard2014c}

Advances in microscopy imaging enable the on-the-fly monitoring of the structural transitions from one morphology to another:
these have been observed experimentally in nanoclusters of a wide size range, transitions being induced by increasing the temperature of the system or through electron beam irradiation.\cite{Koga2004a, Young2010, Wang2012c, Plant2014b}
Solid-solid structural transitions from cuboctahedral (Co) and decahedral (Dh) geometries to the icosahedral (Ih) one via the Diamond-Square-Diamond (DSD)\cite{Mackay1962b, Lipscomb1966b, Wales1996a} mechanism have been reproduced in simulations of mono- and bi- metallic magic-size nanoclusters.\cite{Li2002, Zhang2006, Chen2008a, Cheng2013d, Lan2014, Gould2015, Pavan2015}
The DSD is a direct geometric interconversion mechanism happening via a collective screw dislocation motion.
In principle, it can always connect the aforementioned geometries, irregardless of the number of atoms in the cluster and also when metallic atoms are treated as hard spheres.
On the other hand the energetic cost of cooperative collective rearrangements is expected to increase with the size of the system.\cite{Trygubenko2004}
Yet, previous study did not sistematically addressed whether or not rearrangements through the DSD mechanism are always accessible and energetically favourable.

Here, we report on room temperature structural transitions in noble and quasi-noble metals nanoclusters with Icosahedral magic size of 55, 147, 309 and 561 atoms.
We identify intriguing size and composition dependent effects on the rearrangement towards more energetically favourable structures:
the DSD mechanism results hindered or suppressed after a certain size.
Material inter-atomic potential and surface properties assuming a significant role in determining the critical size of the transition to less collective rearrangement motions.

\begin{table*}[t!]
\centering
\caption{Per each metallic species an cluster size, we report the potential stickyness as given in Ref.\cite{Baletto2002f}~; the first geometrical basin(s) visited when starting from the Dh or Co shape -when two geometries are reported the first refers to the  itMD sampling, and the second to MetaD- highest CP index for the sampled rearrangements (described in the official publication);  $T_{\rightarrow Ih}$, see text; $\Delta F$ (eV) to escape/return from/to the Dh or Co basin.}
\label{Tab:table}
\begin{tabular}{ p{1cm} p{1cm} | p{1.5cm} p{1.5cm} p{1.5cm} p{1.5cm} | p{1.5cm} p{1.5cm} p{1.5cm} p{1.5cm}}
\hline
\hline
System & $\sigma$ & Dh & & & & Co & & & \\
\hline
 & & transition & CP$_{max}$ & T & $\Delta F$ & transition & CP$_{max}$ & T & $\Delta F$ \\
\hline
Ni$_{55}$  & 12.9 & Ih       & 1    &   $<$ 0.02 & 0.05(1.25) & Ih   & 1 & $<$ 0.1 &  $<$0.05               \\
Ni$_{147}$ &      & Ih       & 0.78 &   0.41     & 0.2        & Ih   & 0.88 &  0.38   &  0.1(3.5)            \\
Ni$_{309}$ &      & mDh/Ih   & 0.43 &    $>$ 1   & 0.6   & fcc / Ih  & 0.45 & $>$ 1   &  0.3                 \\
Ni$_{561}$ &      & mDh      & 0.16 &    $>$ 1   & 1.3   & fcc       & 0.13 & $>$ 1   &  0.7                 \\
\hline
Pd$_{55}$  & 20.3 & Ih       & 0.79 &   0.11    & 0.05(1.4)   & Ih   & 0.84 &   0.23  &  0.05                \\
Pd$_{147}$ &      & Ih       & 0.75 &   0.55    & 0.9         & Ih   & 0.83 &   0.60  &  0.5(3.0)            \\
Pd$_{309}$ &      & mDh/Ih   & 0.47 &    $>$ 1  & 3.8    & fcc / Ih  & 0.47 &  $>$ 1  &  3.7                 \\
Pd$_{561}$ &      & mDh      & 0.11 &    $>$ 1  &  /     & fcc       & 0.15 &  $>$ 1  &   /                   \\
\hline
Pt$_{55}$  & 20.6 & Ih       & 0.82 &   0.32    & 0.3(1.3)    & Ih   & 0.88 &   0.39  &   0.25               \\
Pt$_{147}$ &      & Ih       & 0.78 &   0.53    & 1.9         & Ih   & 0.88 &   0.55  &  1.7                 \\
Pt$_{309}$ &      & mDh/Ih   & 0.45 &    $>$ 1  & 7.4    & fcc       & 0.44 &   $>$ 1 &   /                   \\
Pt$_{561}$ &      & mDh      & 0.16 &    $>$ 1  &  /     & fcc       & 0.14 &   $>$ 1 &   /                   \\
\hline
Cu$_{55}$  & 12.1 & Ih       & 1 &  $<$ 0.02 & <0.05(1.3)  & Ih   & 1 &  $<$ 0.02 & $<$0.05               \\
Cu$_{147}$ &      & Ih       & 1 &  $<$ 0.02 &  0.2        & Ih   & 1 &  $<$ 0.02 &  0.1                \\
Cu$_{309}$ &      & Ih       & 0.46 &   0.24    &  0.6   & Ih        & 1 &  $<$ 0.02 &  0.3                \\
Cu$_{561}$ &      & Ih       & 0.47 &   0.61    &  1.3   & Ih        & 0.48 &  0.33     &  0.7                \\
\hline
Ag$_{55}$  & 17.3 & Ih       & 0.88 &  0.17     &  0.05(1.4)  & Ih   & 0.83 &  $<$ 0.02 & $<$0.05               \\
Ag$_{147}$ &      & Ih       & 0.80 &  0.50     &  0.6(4.3)   & Ih   & 0.81 &  0.47     &  0.4(3.2)            \\
Ag$_{309}$ &      & mDh/Ih   & 0.41 &    $>$ 1  &  3.7   & fcc / Ih  & 0.42 &     $>$ 1 &  2.5                \\
Ag$_{561}$ &      & mDh      & 0.13 &    $>$ 1  &   /    & Ih        & 0.16 &     $>$ 1 &   /                  \\
\hline
Au$_{55}$  & 20.7 & Ih       & 0.85 &  0.43     & 0.05(0.9)   & Ih   & 0.86 &   0.40     & $<$0.05              \\
Au$_{147}$ &      & mDh/Ih   & 0.32 &    $>$ 1  &  0.5        & fcc / Ih  & 0.42 &     $>$ 1  & 0.5                \\
Au$_{309}$ &      & mDh      & 0.17 &    $>$ 1  &   /    & fcc       & 0.31 &    $>$ 1   &  /                 \\
Au$_{561}$ &      & mDh      & 0.14 &    $>$ 1  &   /    & fcc       & 0.13 &     $>$ 1  &  /                 \\
\hline
\hline
\end{tabular}
\end{table*}

To characterise rare events such as structural transition in metallic nanoparticles we use two synergic metodologies: Molecular Dynamics simulations coupled with Metadynamics \cite{Laio2008a} (MetaD) at 300 K
and six averaged runs of increasing temperature Molecular Dynamics (itMD) where the temperature is raised by 1 0K every 1 ns starting from 100 K up to melting. 
A velocity Verlet algorithm is used to solve Newton's equations of motion. 
A time step of 5 fs is considered and the temperature, fixed around 300K, is controlled via an Andersen thermostat.
Metal-metal interactions are modelled within the second moment approximation of the tight binding theory according to the Rosato-Guillope-Legrande\cite{Rosato2006} formulation and parametrizations found in the literature\cite{a,b,c}.

\begin{figure*}[t!]
\includegraphics[width=18cm]{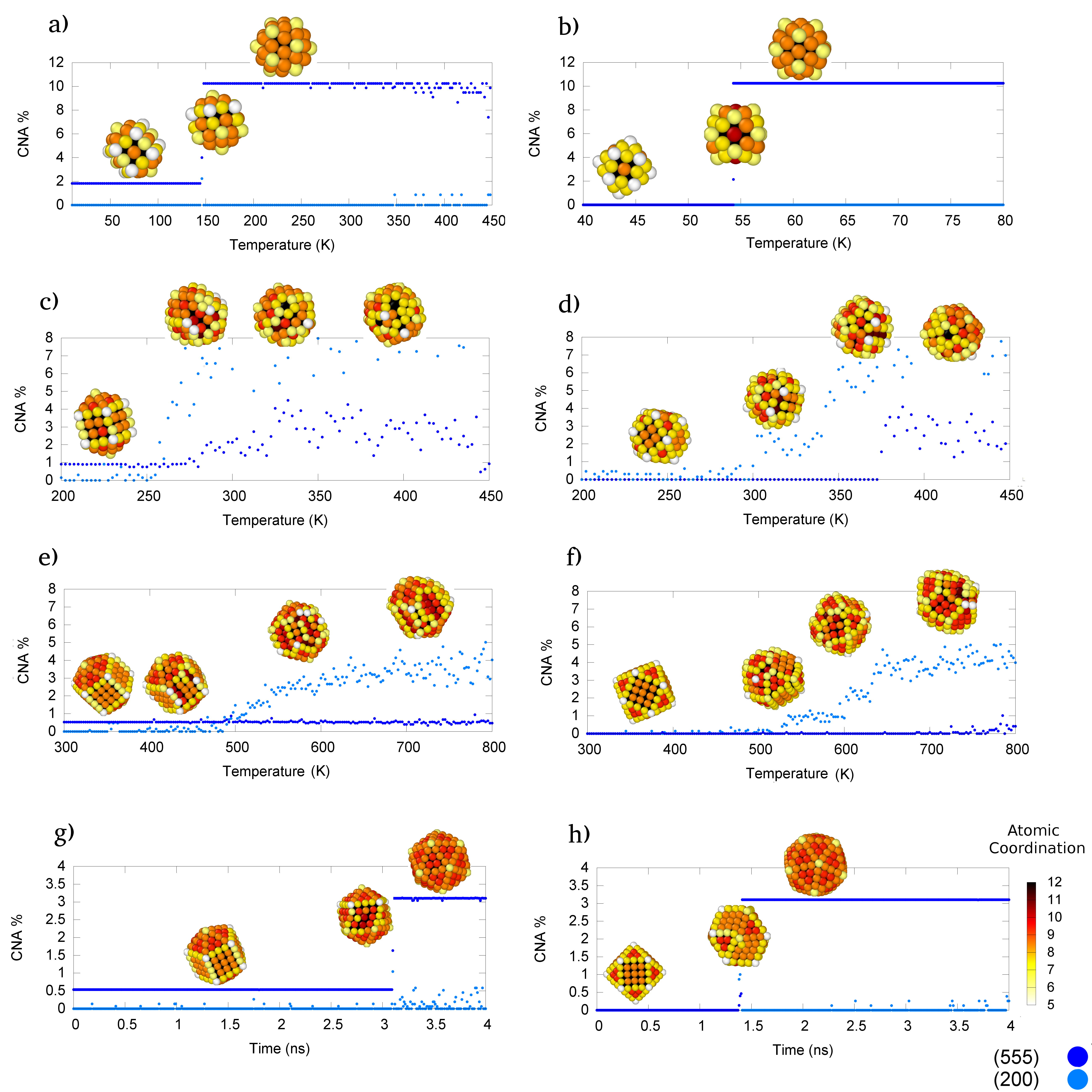}
\caption{(555)\% (in blue) and (200)\% (in cyan) CNA signature evolution for paradigmatic structural transition. Snapshot of the initial, final and representative intermediate structures are reported, atoms are coloured with respect to their coordination (CN). Rearrangement mechanism transition are divided according to 8 different families. a) quintuple Square-Diamond connecting Dh to Ih. b) sixtuple Square-Diamond connecting Co to Ih. c) Dh to defected Ih via surface peeling and DSD-like rearrangement. d) Co to defected Co via (111) facet rearrangement, to defect Ih via surface reconstruction. e) Dh to defected Dh / quasi Marks Dh via surface peeling. f) Co to defected Co via dislocation motion and surface peeling. g) Dh to Ih via layer by layer screw mechanism. h) Co to Ih via layer by layer screw mechanism.}
\label{Fig:figure}
\end{figure*}

Metadynamics (MetaD) coarse grains the dynamics of the system in an order parameter space, also known as collective variable space, where an history dependent potential is built to enhance a wide exploration of the conformational space of the system.\cite{Laio2008a}
Recently, we proposed a set of collective variables to investigate solid-solid transitions between closed-shell nanoclusters of large size with a relatively small computational expanses.
It consists of two window functions set on characteristic distances of the pair distance distribution function of the nanoalloy which promote the sliding and rotation of (111) planes but also single atom rearrangements leading to entropically favourable defected structures.\cite{Pavan2015}
The two window function are set respectively at 1.345 (SFN) and at 2.2 -for 55 atoms clusters- or 3.4  -for 147, 309, 561 atoms clusters- (MPDD) times the lattice distance parameter.
Metadynamics history dependent potential evolves through Gaussians deposited every 10 ps, their height and width are set up homogeneously for all the investigated systems, according to their size: 
55 atoms cluster are perturbed via gaussian 0.1 eV high and with a width of 15 and 10 in the SFN and MPDD collective variable space. The height of the gaussian is increased by 0.1 for each additional layer present in the cluster, up to 0.4 eV in the case of 561 atoms clusters. Similarly gaussians width along SFN and MPDD increase by 20 and 30 for each additional layer.

Rearrangement mechanisms dependence on the size and metal emerges from our $\it{in-silico}$ study.
Table \ref{Tab:table} resumes the energetics of relaxed geometries, the martensinic temperature for structural rearrangement into Ih (if witnessed), and whether or not MetaD samples transitions from the initial structure on the right side of the arrow to the motif on the left. 
For diffusive exploration of two morphological basins, the free energy barriers (FEBs) evaluated according to \cite{Laio2002} is reported.
In the case of non diffusive regime, an heuristic overestimate of the FEB is evaluated as the highest point in the energy landscape reconstructed from Metadynamics potential when the initial Dh or Co minima are left.  
The observed rearrangement mechanisms are categorized according to the architectures families shown in Figure \ref{Fig:figure}.
We highlight that the martensitic temperature increases significantly more with size with respect to the melting temperature. 
Further we remark the quite narrow range of temperatures for martensitic transition temperature when normalized with respect to the melting one.
Nanocluster's geometry evolution during each simulation is analysed by Common Neighbour Analysis (CNA).\cite{Honeycutt1987a}
This technique assigns to each pair of nearest neighbours a signature $\it{(r,s,t)}$ depending on the local network of their neighbourhood:
r is the number of common nearest neighbours of the pair, s the number of bonds between the r atoms, and t the length of longest chain between them.
Characteristic signatures percentages discriminate different environments, particularly useful are the (5,5,5) and the (2,0,0) signature:
they respectively characterize pairs along a 5-fold symmetry axis and on (110) surfaces.

Solid-Solid transitions towards the Ih are observed in all the clusters with 55 atoms both through MetaD and itMD simulations. 
Paradigmatic examples of the Dh$\rightarrow$Ih and Co$\rightarrow$Ih transitions during itMD runs are shown in Figure \ref{Fig:figure} panel a) and b) respectively:
it is single step and the putative saddle point configuration displays the diamond facets characteristic of the DSD rearrangement mechanism.
MetaD runs sample diffusively the Ih and Dh basin before exploring a series of low-symmetry structures.
The chosen collective variables result instead unable to probe Ih into Co transition: 
the latter motif is energetically highly unfavourable and Co$\rightarrow$Ih martensinic transitions appear at very low temperature, thus making difficult to reproduce the Ih$\rightarrow$Co transition.

Material dependent rearrangement are first observed in clusters of 147 atoms: during itMD, Co$\rightarrow$Ih and Dh$\rightarrow$Ih martensic transitions via DSD mechanism are observed in all systems except for gold nanoclusters.
In Au$_{147}$ with an initial Co geometries, Co $\rightarrow$ defected Co transition take place: atoms in one of the triangular (111) facet rearrange into an fcc island with a (111) interface or (110) re-entrance with the neighbouring (100) facets. 
More defects and fcc islands are formed with increasing temperature.
A reconstruction from a low symmetry structure towards a defected Ih with a rosette\cite{Apra2004} defect at one or two vertexes is occasionally witnessed prior to melting.
For a Au$_{147}$ starting Dh, Dh $\rightarrow$ defected Ih transition occurs via a DSD-like mechanism coupled to single atom surface diffusion and rosette defects formation.
Figure \ref{Fig:figure} panel c) and d) show the Au$_{147}$ cluster structural transition during paradigmatic itMD runs.

MetaD, imposing a collective bias, samples rearrangement towards Ih via DSD mechanism in all systems. 
Ag$_{147}$, Pd$_{147}$ and Ni$_{147}$ runs display a diffusive exploration of the Co and Ih basin.
Non diffusive sampling is observed for Au$_{147}$ and Pt$_{147}$: 
these systems displayed only Ih $\rightarrow$ defected Ih or rosette Ih\cite{Apra2004} transitions.
We remark that rosette Ih motifs prevent the formation of a perfect Ih and hinders the rearrangement via DSD mechanism:
because of the low symmetry of these architecture, a single rotational axis is available to rearrange via DSD towards a rosette Dh.
This structure, however, has never been identified in previous studies as a favourable one.
Further, the entropic contributions of clusters presenting structural defects are often higher with respect to the one of closed-packed geometries,\cite{Apra2004} hence they may result competitive with Dh and Co and hide these two conformational basins during MetaD sampling.

While at smaller size the martensinic transitions towards the Ih have been detected consistently by itMD, they have not ever been sampled in clusters of 309 atoms, apart for Cu$_{309}$,
in agreement with what observed in previous studies employing similar temperature increase rate.\cite{Chen2011a, Lan2014}
Co structures rearrange one or more triangular (111) facets into fcc islands.
Dh reorganize into structure presenting defects and Marks' like re-entrances\cite{Marks1994b} formed by surface peeling.\cite{Niiyama2011a}
The clusters preserve their overall geometry as shown by their roughly constant (5,5,5) CNA signature. More and more defects can be appreciated with temperatures as high as 600K
 
Snapshots and structural characterization during itMD are reported in Figure \ref{Tab:table} panel e) and f).
Co $\rightarrow$ defected Co and Dh $\rightarrow$ quasi Marks transitions via the previously described mechanisms comprising collective screw motions, single atom surface diffusion, and rosette defects formation
are also observed in clusters of 561 atoms, except for the Cu one.

According to MetaD, notwithstanding the competition with surface peeling and defects formation, collective rearrangements via DSD mechanism to and from the Ih basin result accessible for Ag$_{309}$, Pd$_{309}$ and Ni$_{309}$.
Further, we identify another rearrangement pathway connecting Dh and Co geometries to the Ih one during MetaD runs. As illustrated in Figure \ref{Fig:figure} panel g) and h) a layer by layer rotation also enables Dh$\rightarrow$Ih and Co$\rightarrow$Ih transitions.

The comparison between MetaD and itMD runs highlights a fundamental issue in the application of the latter to identify structural rearrangements: the temperature increase can hide possible structural rearrangement mechanism if it is rescaled at a too quick pace.
Yet, it has to be remembered that MetaD enhances collective rearrangement motions, thus they may be promoted also when not favourable with respect to other structural transition mechanisms.

In conclusion, we show how size-effects not only influences melting point and favourable geometries but also the accessible rearrangement mechanisms through Metadynamics and increasing temperature MD simulation.
We observe a size-dependent transition from collective motions, such as the Diamond-Square-Diamond,\cite{Lipscomb1966b} to less collective ones as layer-by-layer dislocations or single atomic ones, such as surface peeling\cite{Niiyama2011a} and defects formation,\cite{Apra2004}, its onset being material dependent.
Rearrangement via DSD mechanism result already hindered if not forbidden in Au clusters of 147 atoms.
Vice versa the critical size where less cooperative mechanisms are observed appears to be 309 for Pt clusters and 561 for Ni, Ag and Pd clusters.
In the case of Cu clusters, it is above the sizes here considered.
We observe a correspondence between structural crossover sizes and mechanism crossover ones.
Au and Pt favour Marks-like re-entrances geometries already for clusters larger than 100 atoms.
Ag, Ni and Pd instead prefer Marks-Dh geometries after 300 atoms circa.
Further, Marks-like structure result favourable only in Cu clusters larger that 1000 atoms.\cite{Baletto2002f}
Thus the size influence on the rearrangements of monometallic clusters can be rationalized in terms of nanocluster material stickyness of the inter-atomic potential.

\bibliography{bib}
\bibliographystyle{jcp}

\end{document}